\documentclass[aps,prl,reprint,superscriptaddress,nofootinbib]{revtex4-2}

\usepackage{amsmath}
\usepackage{amssymb}
\usepackage{graphicx}
\usepackage{dcolumn}
\usepackage{bm}
\usepackage{braket}
\usepackage{mathrsfs}
\usepackage{slashed}

\begin{document}


\title{
Nucleon thermalization hindered by isospin symmetry: Violation of eigenstate thermalization hypothesis in atomic nuclei 
}

\author{Dong Bai}
\email{dbai@hhu.edu.cn}
\affiliation{College of Mechanics and Engineering Science, Hohai University, Nanjing 211100, China
}%
\affiliation{Shanghai Research Center for Theoretical Nuclear Physics, NSFC and Fudan University, Shanghai 200438, China}

\author{Zhongzhou Ren}
\email{zren@tongji.edu.cn}
\affiliation{School of Physics Science and Engineering, Tongji University, Shanghai 200092, China}%
\affiliation{Key Laboratory of Advanced Micro-Structure Materials, Ministry of Education, Shanghai 200092, China}


\begin{abstract}

Bohr's compound nucleus theory is one of the most important models in nuclear physics, with far-reaching applications in nuclear science and technology. This model generally assumes that the participating nucleons attain a thermal equilibrium characterized by the microcanonical ensemble before subsequent decays. However, from a theoretical viewpoint, it remains uncertain whether this assumption is universally valid. In this Letter, we critically examine this longstanding assumption through the lens of the eigenstate thermalization hypothesis (ETH), a cornerstone of the modern quantum thermalization theory. Utilizing the time-dependent configuration interaction shell model, it is found that, in certain cases, the long-time averages of nucleon occupation numbers can exhibit significant deviations from the microcanonical ensemble averages, in contrast to the conventional expectation. We attribute this discrepancy primarily to the violation of the ETH in the presence of isospin symmetry and discover that incorporating a substantial isospin-breaking term into the shell-model Hamiltonian can effectively restore the nucleon thermalization.

\end{abstract}

\maketitle


\paragraph*{Introduction.}
In 1936, Bohr postulated that after the projectile is fully absorbed by the target in a nuclear reaction, the participating nucleons can reach a temporary thermal equilibrium state following a series of nucleon-nucleon collisions mediated by nuclear forces \cite{Bohr:1936zz}. This intermediate state, known as the compound nucleus, eventually decays into more stable products through various channels. Due to its thermal nature, the decay process of the compound nucleus is predicted to be independent of its formation process, a result also known as the Bohr independence hypothesis. As a representative mechanism for nuclear reactions, the compound nucleus reaction is extensively used in nuclear science and technology, spanning from uncovering the chaotic aspects of nuclear dynamics \cite{Zelevinsky:1996,Weidenmuller:2008vb,Mitchell:2010um,Papenbrock:2007mq}, modeling the $r$-process in nuclear astrophysics \cite{Horowitz:2018ndv,Cowan:2019pkx}, to improving nuclear data evaluation that supports real-world nuclear applications \cite{Bernstein:2019nqq}. 

Testing Bohr's compound nucleus theory has been a topic of interest in experimental nuclear physics for a long time. 
In 1950, Ghoshal produced the ``same'' compound nucleus ${}^{64}\text{Zn}^*$ through two different entrance channels and observed qualitative agreement with the Bohr independence hypothesis in selected exit channels \cite{Ghoshal:1950zz}.
In the following years, several groups improved upon Ghoshal's experimental scheme and provided valuable insights into the impacts of conservation laws on the Bohr independence hypothesis \cite{John:1956zz,Benveniste:1968zz,Fluss:1969gi,Montgomery:1970nk,Vaz:1972zza,Wiley:1973msg,Go:1973zz}.
Despite these achievements, a more fundamental problem remains largely unexplored: Whether nuclear forces can truly drive the participating nucleons of a compound nucleus reaction to the stage of thermal equilibrium. 
By ``thermal equilibrium'', we mean that the relevant physical state is well described by the standard equilibrium statistical ensembles, a usage consistent with the common convention \cite{Landau:1980}.
Despite its central role in Bohr's compound nucleus assumption, there has been little direct investigation from a theoretical perspective.
On the one hand, it is exceedingly challenging to describe the realistic compound nucleus formation process at the level of nucleons in a fully quantum mechanical way. The existing models, such as the time-dependent Hartree-Fock (TDHF) method \cite{Simenel:2018euu}, often involve notable simplifications regarding nucleon correlations, and it is hard to evaluate their impacts on nucleon thermalization.
On the other hand, the topic of quantum thermalization has been a source of confusion for decades.
It is only recently that a widely accepted framework has emerged, thanks to intensive investigations into the eigenstate thermalization hypothesis (ETH) \cite{Deutsch:1991,Srednicki:1994,Rigol:2008}, as well as its generalizations and violations \cite{Rigol:2007,Murthy:2022dao,Fukushima:2023svf,Cipolloni:2023dtl,Abanin:2018yrt} (for comprehensive reviews, see Refs.~\cite{Borgonovi:2016,Ueda:2020,Majidy:2023xhm,Deutsch:2018ulr,DAlessio:2015qtq}).

\begin{figure*}

\centering

  \includegraphics[width=0.7\linewidth]{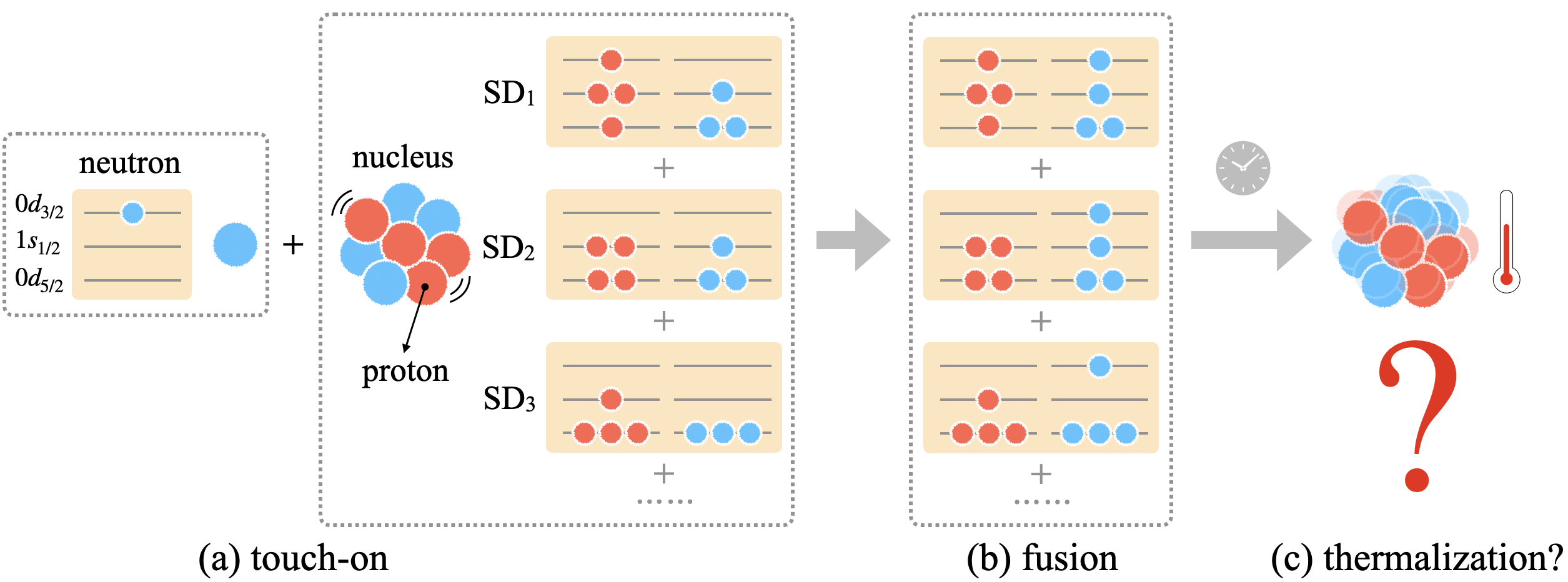}  

  \caption{\emph{A simplified model for the compound nucleus formation in a neutron-capture reaction.} (a) A projectile neutron, depicted in blue and prepared to be in a specific single-particle state, touches upon the target nucleus, which is in an energy eigenstate composed of Slater determinants (SDs, represented by orange blocks filled with nucleons). At this point, no interaction exists between the neutron and the target nucleus.   
(b) The neutron then fuses with the target nucleus, establishing the initial state for the subsequent time evolution. 
Since there is no neutron-nucleus interaction, the neutron simply occupies empty neutron orbits within the target nucleus.  
(c) After the neutron-nucleus interaction is activated, the compound system starts to evolve over time according to the full Hamiltonian. After a significant period, a steady state arises, and a longstanding but theoretically unverified assumption posits that this state can be described by the microcanonical ensemble. 
  }
  \label{pic:fusion}

\end{figure*}

In this Letter, we give a comprehensive exploration of nucleon thermalization driven by nuclear forces in the light of the ETH,
which has been found to be valid in a number of condensed matter and ultracold atomic systems.
Very recently, it has been shown that the ETH is obeyed by the (2+1)-dimensional SU(2) gauge theory,
which is a close relative to quantum chromodynamics, the fundamental theory underlying nuclear physics \cite{Yao:2023pht,Ebner:2023ixq}.
In Ref.\ \cite{Mueller:2024mmk}, the thermalization dynamics of a (2+1)-dimensional lattice gauge theory is explored using the state-of-the-art quantum computing technique.
In spite of these achievements, it is fair to say that the applicability of the ETH in atomic nuclei is largely unexplored,
and it is our goal to bridge this gap.

To clarify the role of nuclear forces, particularly their symmetric properties, in nucleon thermalization, we utilize a simplified model for the formation of a compound nucleus in a neutron-capture reaction. This model is based on a quantum quench in the time-dependent configuration interaction shell model (depicted in Fig.\ \ref{pic:fusion}). In this approach, nucleons are confined to the standard shell-model space (such as the $sd$- and $pf$-shells), preventing any escape into the continuum. This confinement offers three key advantages. 
First, it excludes competitive reaction mechanisms, such as pre-equilibrium and direct nuclear reactions, where nucleons may escape from the compound system before thermal equilibrium can be established. 
Second, by solving the configuration interaction shell model using exact diagonalization techniques \cite{Johnson:2018hrx},
our model provides an exact description of nucleon correlations within the model space.
This makes it suitable for exploring the interplay between nucleon thermalization and nuclear forces.
Last, our model aligns with quantum quenches used in condensed matter physics and ultracold atomic physics, thereby allowing a more straightforward comparison between thermalization phenomena in these distinct physical systems.
While previous studies have focused on the thermal properties of nuclear many-body eigenstates \cite{Zelevinsky:1996,Horoi:1994ys,Zelevinsky:2003pi,Zelevinsky:2019iai}, our work concentrates on nucleon thermalization in nuclear dynamics, where quantum states are typically not eigenstates of shell-model Hamiltonians or other conserved charges.

\paragraph*{ETH.}
To simulate the formation of a compound nucleus, we employ a simplified model within the framework of the time-dependent configuration interaction shell model (see Refs.\ \cite{Caurier:2004gf,Otsuka:2018bqq} for reviews of shell models).
Unless otherwise specified, the Hamiltonian used in this model respects the isospin symmetry.
Throughout this Letter, we use the subscripts ``$t$'' and ``$c$''
to distinguish between physical quantities for the target and compound systems.
For instance, the Hamiltonians for the target and compound nuclei are denoted as $\hat{H}_t$ and $\hat{H}_c$.

To initiate the time evolution, we set the initial state as $\ket{\psi(0)}=\hat{c}^\dagger_\alpha\ket{{E}_t,{M}_t}$. This setup simulates the touch-on and fusion between an incident neutron, described by the shell-model single-particle state $\ket{\alpha}=\ket{a,m_\alpha,m_{T\alpha}}=\hat{c}^\dagger_\alpha\ket{0}$, and the target nucleus, which resides a specific eigenstate $\ket{{E}_t,{M}_t}$.
Here, 
$a$ denotes the single-particle orbit within the shell model,
$m_\alpha$ and $m_{T\alpha}$ represent the magnetic quantum number and the $z$-component of the isospin for the single-particle state,
$\hat{c}^\dagger_\alpha$ is the creation operator corresponding to $\ket{\alpha}$,
while ${E}_t$ and ${M}_t$ are the eigenenergy and the magnetic quantum number of the target nucleus.
Additionally, $\ket{E_t,M_t}$ implicitly carries other good quantum numbers: $J_t$, $T_t$, and $M_{Tt}$, which are associated with $\hat{J}_t^2$, $\hat{T}_t^2$,
and $\hat{T}_{zt}$.
It is evident that $\ket{\psi(0)}$ is an eigenstate of $\hat{J}_{zc}$ and $\hat{T}_{zc}$ for the compound system, with the corresponding quantum numbers given by ${M}_c=m_\alpha+M_t$ and $M_{Tc}=m_{T\alpha}+M_{Tt}$.
However, it is typically not an eigenstate of $\hat{H}_c$, $\hat{J}_c^2$, and $\hat{T}_c^2$.

\begin{table}[th]
  \centering
      \begin{tabular}{llcc}
    \hline\hline\\[-2ex]
   & & \multicolumn{1}{c}{\ \ $n+{}^{49}\text{Ca}^*$\ \ } &  \multicolumn{1}{c}{\ \ $n+{}^{23}\text{Mg}^*$\ \ } \\
    \hline
general & model space & $pf$-shell & $sd$-shell \\
information\ \ &  interaction & KB3G \cite{Poves:2000nw} & USDB \cite{Brown:2006gx} \\
& active nucleons & $10$ & $8$ \\
& space dimension\ \  & 17276 & 28503 \\
    \hline
  $n$ & orbit & $0f_{7/2}$ & $0d_{3/2}$  \\
 & $m_\alpha$ & $-1/2$  & $-1/2$ \\
  & $m_{T\alpha}$ & $1/2$  & $1/2$ \\
    \hline
 nucleus & $i$th eigenstate & 10000 & 9851 \\
&    $J_t$ & $13/2$ & $11/2$ \\
 &   $T_t$ & $9/2$ & $7/2$ \\
 &   $M_t$ & $1/2$ & $1/2$ \\
 &   $M_{Tt}$ & $9/2$ & $7/2$ \\
    \hline\hline
    \end{tabular}%
  \caption{\emph{The general information for modeling the two fusion processes and the initial states.}
  Rows 2--5 list the shell-model spaces, nuclear interactions, numbers of active nucleons, and Hilbert-space dimensions for these two processes.
  Here, both the KB3G and USDB interactions obey the isospin symmetry.
  Rows 6--8 list the quantum numbers $(a,m_\alpha,m_{T\alpha})$ for the incident nucleons.
  Row 9 specifies the selected eigenstates for the target nuclei ${}^{49}\text{Ca}$ and ${}^{23}\text{Mg}$,
  while Rows 10--13 list the additional quantum numbers $(J_t,T_t,M_T,M_{Tt})$.
  }
  \label{tab:Initial-State}%
\end{table}%

\begin{figure*}

\centering

  \includegraphics[width=\linewidth]{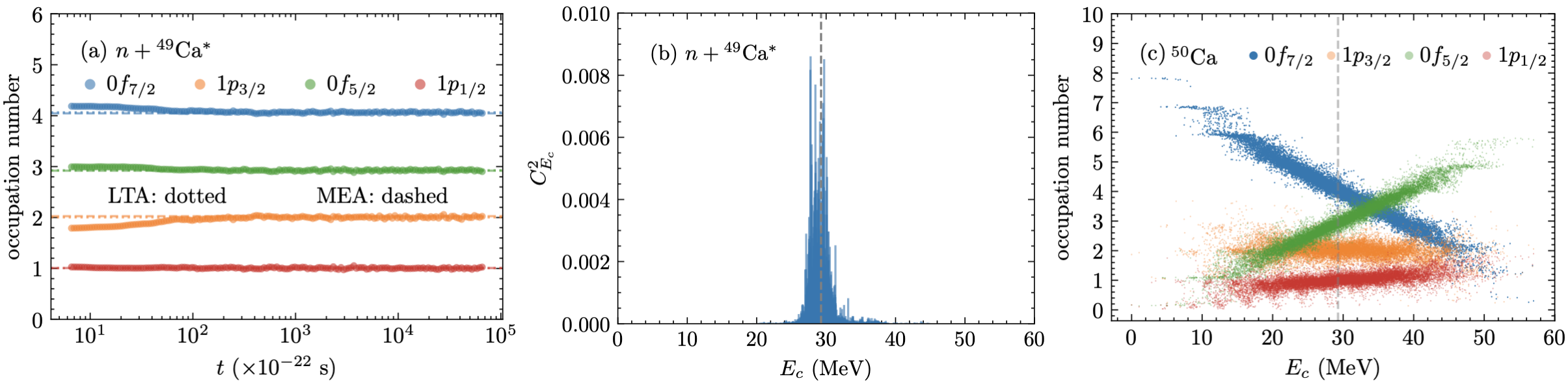}  

  \caption{\emph{The emergence of nucleon thermalization.} 
  (a) The time evolution of the occupation number $n_a(t)=\sum_{m_\alpha}\sum_{m_{T\alpha}}$ $\braket{\psi(t)|\hat{c}^\dagger_\alpha \hat{c}_\alpha|\psi(t)}$ is shown for the four single-particle orbits 
  $0f_{7/2}$ (the blue points), $1p_{3/2}$ (the orange points), $0f_{5/2}$ (the green points), and $1p_{1/2}$ (the red points) in the $n+{}^{49}\text{Ca}^*$ fusion process.
  For the microcanonical ensemble calculations, we take $\Delta E_c=0.05$ MeV.
  The long-time averages (LTA, the dotted lines) of $n_a(t)$ agree well with the microcanonical ensemble averages (MEA, the dashed lines),
  indicating the emergence of nucleon thermalization. 
  (b) The distribution of the coefficient $C^2_{E_c}$ is shown for the $n+{}^{49}\text{Ca}^*$ fusion process, where $C^2_{E_c}$ is peaked around the average energy $\overline{E}_c$ (the gray dashed line).
  (c) The eigenstate expectation value $n_a(E_c)=\sum_{m_\alpha}\sum_{m_{T\alpha}}\braket{{E}_c,{M}_c|\hat{c}^\dagger_\alpha \hat{c}_\alpha|{E}_c,{M}_c}$ is shown for the four orbits. For each orbit, $n_a(E_c)$, as a function of $E_c$, resembles a smooth curve consistent with the ETH.  
  }
  \label{pic:Ca50-LTE}

\end{figure*}

Expanding $\ket{\psi(0)}$ in terms of the eigenstates $\ket{{E}_c,{M}_c}$ of the compound system, we have $\ket{\psi(0)}=\sum_{E_c}C_{E_c}\ket{{E}_c,{M}_c}$. Here, $M_c$ is fixed to the magnetic quantum number of $\ket{\psi(0)}$,  
$\sum_{E_c}$ represents the summation over $E_c$, 
and $C_{E_c}$ is the real-valued coefficient normalized by $\sum_{E_c} C_{E_c}^2=1$. 
The subsequent time evolution is then given by
$\ket{\psi(t)}=\exp(-i\hat{H}_ct)\ket{\psi(0)}=\sum_{E_c}$ $C_{E_c}\exp(-iE_ct)\ket{{E}_c,{M}_c}$. 
Thanks to the large hierarchy between nuclear and macroscopic scales,
the formation process of a compound nucleus can be viewed safely as an isolated process. 
If $\ket{\psi(t)}$ reaches thermal equilibrium at late time,
the long-time average (LTA) of a few-body operator $\hat{\mathcal{O}}$ should be approximately equal to its microcanonical ensemble average (MEA) at $\overline{E}_c=\braket{\psi(0)|\hat{H}_c|\psi(0)}$:
 \begin{equation}
\braket{\hat{\mathcal{O}}}_\text{LTA}\approx\braket{\hat{\mathcal{O}}}_{\text{MEA},\overline{E}_c},\label{QT}
\end{equation}
where 
$\braket{\hat{\mathcal{O}}}_\text{LTA}=\lim_{\mathcal{T}\to\infty}\frac{1}{\mathcal{T}}\int_0^{\mathcal{T}}\mathrm{d}t'\braket{\psi(t')|\hat{\mathcal{O}}|\psi(t')}=\sum_{E_c} C_{E_c}^2\braket{{E}_c,{M}_c|\hat{\mathcal{O}}|{E}_c,{M}_c}$ and 
$\braket{\hat{\mathcal{O}}}_{\text{MEA},\overline{E}_c}=\frac{1}{\mathcal{N}_{\overline{E}_c,\Delta E_c}}$ $\sum_{|E_c-\overline{E}_c|<\Delta E_c}\braket{{E}_c,{M}_c|\hat{\mathcal{O}}|{E}_c,{M}_c}$.
In the definition of $\braket{\hat{\mathcal{O}}}_{\text{MEA},\overline{E}_c}$, $M_c$ is not a free variable but fixed to the magnetic quantum number of $\ket{\psi(0)}$, while
$\mathcal{N}_{\overline{E}_c,\Delta E_c}=\sum_{|E_c-\overline{E}_c|<\Delta E_c}1$ is the number of eigenstates within the energy shell $|E_c-\overline{E}_c|<\Delta E_c$, regardless of $J_c$ and $T_c$.
%
The ETH is aimed at explaining when and why thermal equilibrium, as stated by Eq.\ \eqref{QT}, can emerge from an isolated quantum system.
The key assumption of the ETH is that 
an eigenstate of quantum many-body systems has already been self-thermalized secretly, and
$\braket{{E}_c,{M}_c|\hat{\mathcal{O}}|{E}_c,{M}_c}$ is approximately a smooth function of $E_c$, resembling a curve in the $E_c$-$\braket{{E}_c,{M}_c|\hat{\mathcal{O}}|{E}_c,{M}_c}$ plane,
with the deviations suppressed by the Hilbert-space dimension.
This naturally gives rise to the relation $\braket{{E}_c,{M}_c|\hat{\mathcal{O}}|{E}_c,{M}_c}\approx\braket{\hat{\mathcal{O}}}_{\text{MEA},E_c}$, which, along with the observation that $C^2_{E_c}$ often exhibits a clustered distribution centered around $\overline{E}_c$, leads to Eq.\ \eqref{QT} \cite{Deutsch:2018ulr,DAlessio:2015qtq}.

\begin{figure*}

\centering

  \includegraphics[width=\linewidth]{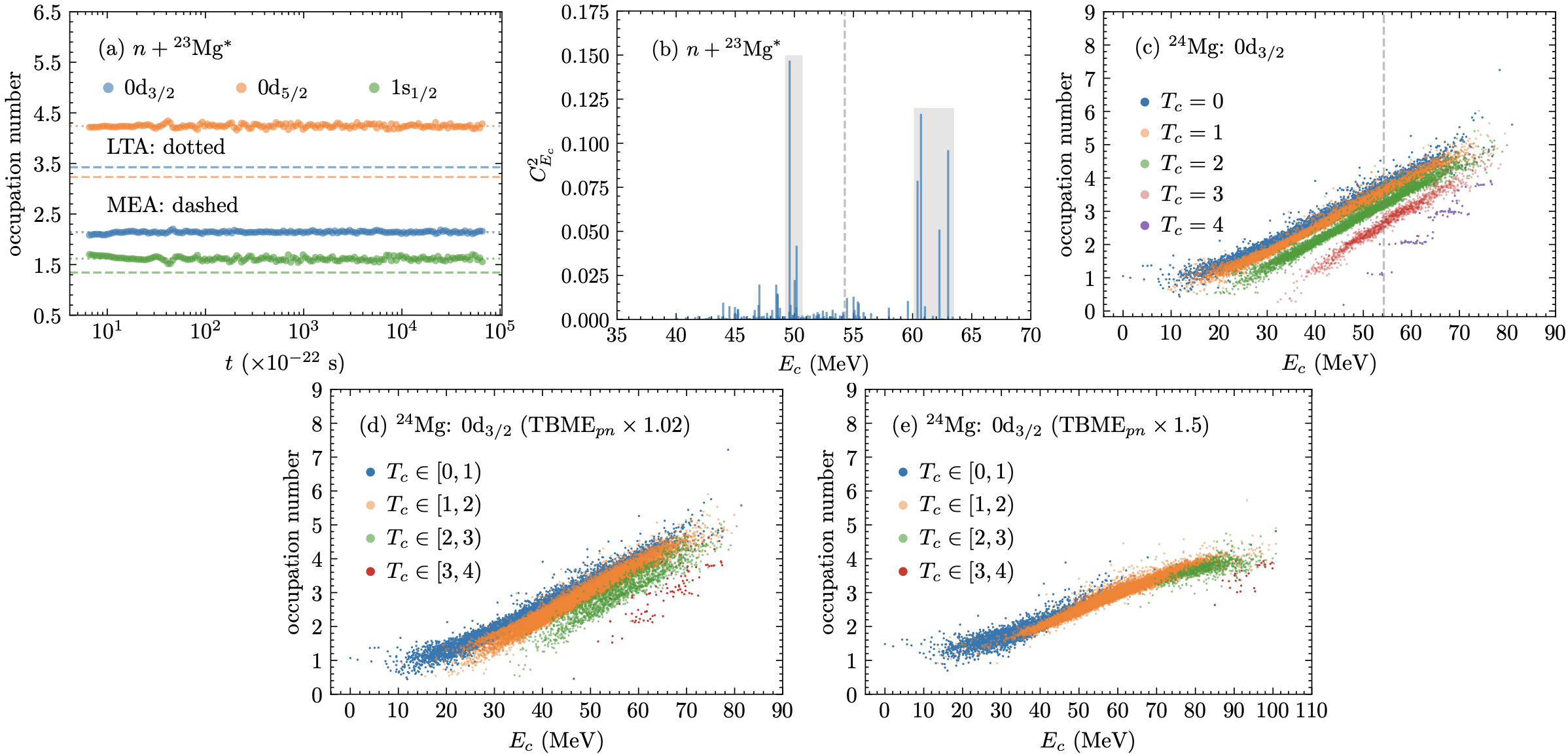}  

  \caption{\emph{The hindrance of nucleon thermalization.}
  (a) The time evolution of the occupation number $n_a(t)$ is shown for the three single-particle orbits 
  $0d_{3/2}$ (the blue points), $1d_{5/2}$ (the orange points), and $1s_{1/2}$ (the green points) in the $n+{}^{23}\text{Mg}^*$ fusion process.
  To calculate the MEA values, we take $\Delta E_c=0.1$ MeV.
  The LTAs (the dotted lines) of $n_a(t)$ deviate significantly with the MEAs (the dashed lines),
  indicating that the nucleon thermalization gets hindered in this case.
  (b) The coefficient $C^2_{E_c}$ is shown as a function of the eigenenergy $E_c$,
  where two peaks appear on both sides of $\overline{E}_c$ (the gray dashed line).
  (c) The eigenstate expectation value $n_a(E_c)$ is shown for the $0d_{3/2}$ orbit as a function of the eigenenergy $E_c$ of ${}^{24}$Mg,
  with data points in different colors according to the $T_c$ value of $\ket{E_c,M_c}$.  
  (d) $n_a(E_c)$ is shown for the $0d_{3/2}$ orbit in the presence of isospin symmetry breaking, where the two-body matrix elements between protons and neutrons ($\text{TBME}_{pn}$s) are rescaled by a factor of 1.02 to break the isospin symmetry. The numerical results are colored according to the average isospin $T_c$ of $\ket{E_c,M_c}$. 
  (e) The same as Fig.\ \ref{pic:Mg24-LTE}(d) except that the $\text{TBME}_{pn}$s are rescaled by a factor of 1.5. As the scaling factor is increased from 1.02 to 1.5, $n_a(E_c)$ is transformed into a smooth function of $E_c$, resembling a single curve.
   }
  \label{pic:Mg24-LTE}

\end{figure*}

\paragraph*{Emergence and hindrance of nucleon thermalization.}
We study the formation of compound nuclei in the $n+{}^{49}\text{Ca}^*$ and $n+{}^{23}\text{Mg}^*$ fusion processes,
with the relevant information given in Table \ref{tab:Initial-State}.
The superscript ``$*$'' denotes that the target nuclei are in their excited states.
For simplicity, all energies of the compound system are referenced relative to the ground state.
In the $n +{}^{49}\text{Ca}^*$ fusion process, 
only the quantum states with $T_c=5$ are allowed within the $pf$-shell.
In comparison, in the $n+{}^{23}\text{Mg}^*$ fusion process, the eigenstates with $T_c=0,\cdots,4$ are all allowed for the $sd$-shell, and a general quantum state is a superposition of these eigenstates.
This distinction is crucial for understanding the different fates of nucleon thermalization in these two processes.
Generally, in a nuclear collision involving $Z$ active protons and $N$ active neutrons,
the quantum state incorporates contributions with various isospins spanning from $(N-Z)/2$ to $(N+Z)/2$.

In Fig.\ \ref{pic:Ca50-LTE}(a), we study the time evolution of the occupation number
$n_a(t)$ in the $n+{}^{49}\text{Ca}^*$ fusion process. 
It is found that, for all the four orbits within the $pf$-shell, the LTAs of $n_a(t)$
overlap almost completely with the corresponding MEAs.
For all the four orbits, the relative deviations between the LTA and MEA results are less than 0.5\%,
which means that the nucleon thermalization formulated by Eq.\ \eqref{QT} is valid for the $n+{}^{49}\text{Ca}^*$ fusion process.
In Fig.\ \ref{pic:Ca50-LTE}(b), $C_{E_c}^2$ is found to be distributed narrowly around $\overline{E}_c$.
As shown by Fig.\ \ref{pic:Ca50-LTE}(c),
the eigenstate expectation value $n_a(E_c)$, as a function of $E_c$, resembles a smooth curve for each of the four orbits, which is in good agreement with the ETH.
Along with Fig.\ \ref{pic:Ca50-LTE}(b), this explains the emergence of nucleon thermalization in the $n+{}^{49}\text{Ca}^*$ fusion process.


In Fig.\ \ref{pic:Mg24-LTE}(a), the time evolution of $n_a(t)$ is shown for the $n+{}^{23}\text{Mg}^*$ fusion process.
It is found that the LTA and MEA values are apparently different for the $0d_{3/2}$, $0d_{5/2}$, and $1s_{1/2}$ orbits, with $|\text{MEA}-\text{LTA}/\text{LTA}|\approx60\%$, $24\%$, and $20\%$, which are one order of magnitude larger than those of the $n+{}^{49}\text{Ca}^*$ fusion process.
This implies that the nucleon thermalization is hindered in this case.
In order to understand the origins of this hindrance, 
we analyze the coefficients $C^2_{E_c}$ and the eigenstate expectation value $n_a(E_c)$ with respect to $E_c$. 
As shown in Fig.\ \ref{pic:Mg24-LTE}(b), in contrast to the $n+{}^{49}\text{Ca}^*$ fusion process,
$C^2_{E_c}$ is distributed abnormally in the $n+{}^{23}\text{Mg}^*$ fusion process, clustering into two peaks away from $\overline{E}_c$
instead of one peak centered around $\overline{E}_c$.
Meanwhile,  in Fig.\ \ref{pic:Mg24-LTE}(c), $n_a(E_c)$ is split into five branches with respect to the isospin $T_c$ of the eigenstate $\ket{E_c,M_c}$. 
Although each branch does resemble a smooth curve by itself, 
they cannot be treated as a single curve as a whole.
In other words, the ETH is violated in the $n+{}^{23}\text{Mg}^*$ fusion process, and 
it is straightforward to see that the isospin symmetry plays a vital role in this violation.

\paragraph{Isospin symmetry breaking.}
In the real world, the isospin symmetry, broken by the Coulomb interaction and the isospin non-conserving components of nuclear interactions, is only an approximate symmetry.
It is crucial to explore the impacts of isospin symmetry breaking on nucleon thermalization. 
We adopt the two-body matrix elements of the Coulomb interactions from Ref.\ \cite{Ormand:1989kck} and rescale the two-body matrix elements between protons and neutrons ($\text{TBME}_{pn}$s) to simulate the isospin non-conserving nuclear interactions \cite{Ormand:1989kck,Magilligan:2020bbd}.
 In Fig.\ \ref{pic:Mg24-LTE}(d), the $\text{TBME}_{pn}$s are rescaled by a factor of 1.02, which is chosen to simulate the physical situation of isospin symmetry breaking. 
 Here, $n_a(E_c)$ turns out be similar to the isospin-symmetric case given by Fig.\ \ref{pic:Mg24-LTE}(c), indicating that the ETH is still violated even in the presence of the physical isospin-breaking effect.
As the scaling factor is increased from 1.02 to 1.5, $n_a(E_c)$ is redistributed gradually to resemble a smooth curve
and eventually becomes consistent with the ETH as shown by Fig.\ \ref{pic:Mg24-LTE}(e). 
As a result, a substantial isospin-breaking term restores the nucleon thermalization effectively.

\paragraph*{Conclusions.}
In this Letter, we reexamine Bohr's assumption regarding the emergence of nucleon thermalization in compound nucleus reactions, viewed through the perspective of the ETH.
Our findings indicate that Bohr's assumption is not universally valid, as the isospin symmetry of nuclear Hamiltonians can hinder the nucleon thermalization in specific cases.
We anticipate that this hindrance effect would be more relevant for symmetric nuclear systems compared to asymmetric ones, given that the isospin quantum number typically spans a broader range in symmetric systems.
This result challenges the foundation of the traditional approach to the compound nucleus reactions rooted in statistical mechanics, and underscores the need for a comprehensive reassessment of the accuracy and limitations of existing statistical models,
which can have far-reaching implications for interdisciplinary fields such as nuclear astrophysics, nuclear power, medical imaging, and national security.

 \begin{acknowledgments}
 
\paragraph*{Acknowledgments.} 
D.\ B.\ thanks Guang-Shang Chen, Zhen Fang, Zhen Li, Hao Lu, Yue-Liang Wu, and Shan-Gui Zhou for useful discussions.
D.\ B.\ also acknowledges the hospitality of Institute of Theoretical Physics, Chinese Academy of Sciences during the final stage of this work.
This work is supported by the National Natural Science Foundation of China (Grants No.\ 12375122,
No.\ 12035011, No.\ 11975167,
and No.\ 12147101),
the National Key R\&D Program of China (Contract No.\ 2023YFA1606503),
and the Fundamental Research Funds for the Central Universities (Grant No.\ B230201022 and No.\ B240201048).

\end{acknowledgments}

\end{document}